\documentclass[twocolumn,prl,floatfix,superscriptaddress,nobibnotes,nolongbibliography]{revtex4-2}

\usepackage{amsmath}
\usepackage{amssymb}
\usepackage{graphicx}
\usepackage{hyperref}
\usepackage[normalem]{ulem}

\makeatletter
\renewcommand{\sb}[1]{_{\mathrm{#1}}}

\makeatother

\usepackage{cleveref}
\crefname{figure}{Fig.}{Figs.} 

\usepackage{xcolor}
\usepackage{color}

\hypersetup{
    colorlinks=true,
    linkcolor={blue!80!black},
    citecolor={blue!80!black},
    urlcolor={blue!80!black}
}

\begin{document}

\title{Voltage-tuned anomalous-metal to metal transition in\\hybrid Josephson junction arrays}

\author{S.~Sasmal}
\affiliation{Center for Quantum Devices, Niels Bohr Institute, University of Copenhagen, 2100 Copenhagen, Denmark}
\author{M.~Efthymiou-Tsironi}
\affiliation{Center for Quantum Devices, Niels Bohr Institute, University of Copenhagen, 2100 Copenhagen, Denmark}
\affiliation{Department of Physics, Universitá del Salento, via Monteroni, 73100, Lecce, Italy}
\author{G.~Nagda}
\affiliation{Center for Quantum Devices, Niels Bohr Institute, University of Copenhagen, 2100 Copenhagen, Denmark}
\author{E.~Fugl}
\affiliation{Center for Quantum Devices, Niels Bohr Institute, University of Copenhagen, 2100 Copenhagen, Denmark}
\author{L.~L.~Olsen}
\affiliation{Center for Quantum Devices, Niels Bohr Institute, University of Copenhagen, 2100 Copenhagen, Denmark}
\author{F.~Krizek}
\affiliation{Center for Quantum Devices, Niels Bohr Institute, University of Copenhagen, 2100 Copenhagen, Denmark}
\affiliation{Institute of Physics, Czech Academy of Sciences, 162 00 Prague, Czech Republic}
\author{C.~M.~Marcus}
\affiliation{Center for Quantum Devices, Niels Bohr Institute, University of Copenhagen, 2100 Copenhagen, Denmark}
\affiliation{Materials Science and Engineering, and Department of Physics, University of Washington, Seattle WA 98195}
\author{S.~Vaitiek\.{e}nas}
\affiliation{Center for Quantum Devices, Niels Bohr Institute, University of Copenhagen, 2100 Copenhagen, Denmark}

\date{\today}

\begin{abstract} 

We report voltage-tuned phase transitions in arrays of hybrid semiconductor-superconductor islands arranged in a square lattice. 
A double-layer electrostatic gate geometry enables independent tuning of inter-island coupling and proximity-induced superconductivity. 
This design enables access to the superconductor-insulator, superconductor-metal, and metal-insulator transitions in a single device, revealing critical points and emergent intermediate phases. 
We find that the superconductor-insulator transition is interrupted by an anomalous metallic phase with saturating low-temperature resistivity.
Across gate voltages, this regime extends over three orders of magnitude in resistivity and can be continuously tuned into the conventional metallic phase.
The signature of the anomalous metallic phase is suppressed by magnetic frustration.

\end{abstract}

\maketitle

Two-dimensional systems were long thought to transition directly from a superconducting to an insulating phase with increasing effective disorder~\cite{haviland1989onset, Lee_1990_Critical, gantmakher2010superconductor, Goldman_2010_Superconductor, lin2015superconductivity}.
However, numerous experiments have since reported an intermediate metallic phase characterized by finite, saturating resistivity as temperature, $T$, approaches zero~\cite{kapitulnik2019colloquium, wang2023quantum}. 
This anomalous metallic (AM) phase has been observed in various systems, including thin films \cite{jaeger1989onset, Yazdani_1995_Superconducting, markovic1999superconductor, mason2001true, breznay2017particle}, oxides~\cite{steiner2008approach, bollinger2011superconductor}, crystalline superconductors \cite{li2019anomalous}, patterned islands on graphene \cite{han2014collapse}, perforated high-$T\sb{C}$ superconductors \cite{yang2019intermediate}, and Josephson junction arrays \cite{eley2012approaching, bottcher2018superconducting, bottcher2023dynamical, bottcher2024berezinskii}. 

Several mechanisms have been proposed to explain the AM behavior, including order parameter fluctuations \cite{Feigelman_1998_Quantum, Spivak_2001_Quantum, Spivak_2008_Theory}, emergence of a Bose metal~\cite{Lee_1991_Theory, Kivelson_1992_Global, Phillips_2003_Elusive}, coupling to a dissipative environment~\cite{Mason_1999_Dissipation, Kapitulnik_2001_Effects}, formation of a composite Fermi liquid~\cite{Mulligan_2016_Composite}, residual heating due to insufficient electronic filtering~\cite{tamir2019sensitivity}, and free vortex motion~\cite{vaitiekenas2020anomalous, zhang2021robust}. 
As a result, recent studies have shifted focus from the specifics of the superconductor-insulator transition (SIT) to the broader question of how a metallic state can persist at low $T$ when superconductivity is suppressed by external perturbations \cite{zhang2022anomalous, hsu2024transport, zhang2025gapless}. 
This includes understanding how the anomalous metal relates to a conventional metal, and how its extent depends on the degree of disorder.

Here, we investigate voltage-tuned phase transitions in two-dimensional semiconductor-superconductor hybrid Josephson junction arrays. 
The system features dual electrostatic gates that enable independent control of the coupling between neighboring islands and the coupling between the islands and the surrounding semiconductor. 
This geometry enables experimental access to superconducting, insulating, and metallic states within a single device.
By decreasing the inter-island coupling, we observe an SIT interrupted by an AM phase, with low-temperature saturation resistivity that can be tuned across more than three orders of magnitude around the superconducting resistance quantum, $h/4e^2$.
When tuning the proximity-induced superconductivity independently, the AM phase continuously evolves into a conventional metallic phase.
Finally, we find that frustrating the array by applying a perpendicular magnetic field introduces the $T$ dependence in the resistivity, suppressing the signatures of the AM phase.
We attribute this behavior to thermally activated vortex motion.

\begin{figure}[!t]
    \includegraphics[width=\linewidth]{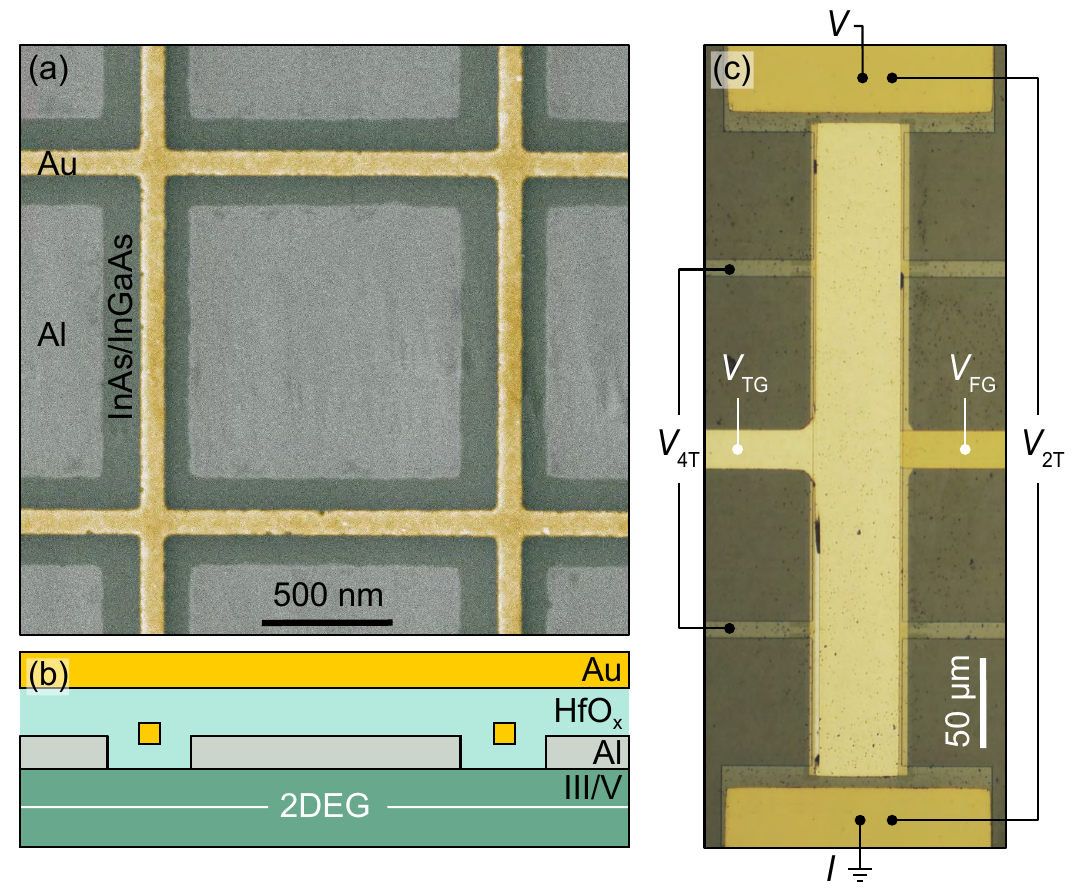}
    \caption{(a) Colorized scanning electron micrograph of a reference device, taken before depositing the global top gate.
    The square Al islands (gray) are patterned on top of the semiconducting heterostructure (green-gray) and are separated by a frame gate (yellow).
    (b) Schematic cross-section of the device illustrating dual gate geometry. The lower frame gate tunes the central part of the junctions, while the global top gate tunes the 2DEG surrounding the islands.
    (c) Optical micrograph of the measured Hall bar device showing measurement setup.}
    \label{fig:1}
\end{figure}

The measured hybrid Josephson junction arrays were fabricated on an InAs two-dimensional electron gas (2DEG) heterostructure proximitized by epitaxial Al, similar to that described in Refs.~\cite{Shabani_2016_Two-dimensional,cheah2023control}.
The Al layer was lithographically patterned into a $255\times31$ array of $1\,\mu$m square islands with 400 nm spacing. 
After covering the structure with a thin HfO$\sb{x}$ dielectric, a 100 nm wide frame gate was metallized between the Al islands to tune their coupling; see \cref{fig:1}(a).
A second HfO$\sb{x}$ layer and a global Ti/Au top gate were added to control the uncovered regions around the islands.
A schematic cross-section of the complete device stack is shown in \cref{fig:1}(b).
An optical image of the final device, overlaid with the measurement setup is shown in \cref{fig:1}(c).
Two additional devices, with different island sizes and spacings, showed similar results. 
In the main text, we focus on data from a representative device.
Supporting data from the other two devices are presented in the Supplemental Material~\cite{Supplement}.
Measurements were performed using standard lock-in techniques in a dilution refrigerator with a three-axis vector magnet and base temperature of 15~mK.
Further details on the growth, sample fabrication, and measurements are provided in the Supplemental Material~\cite{Supplement}.

\begin{figure}[!t]
    \includegraphics[width=\linewidth]{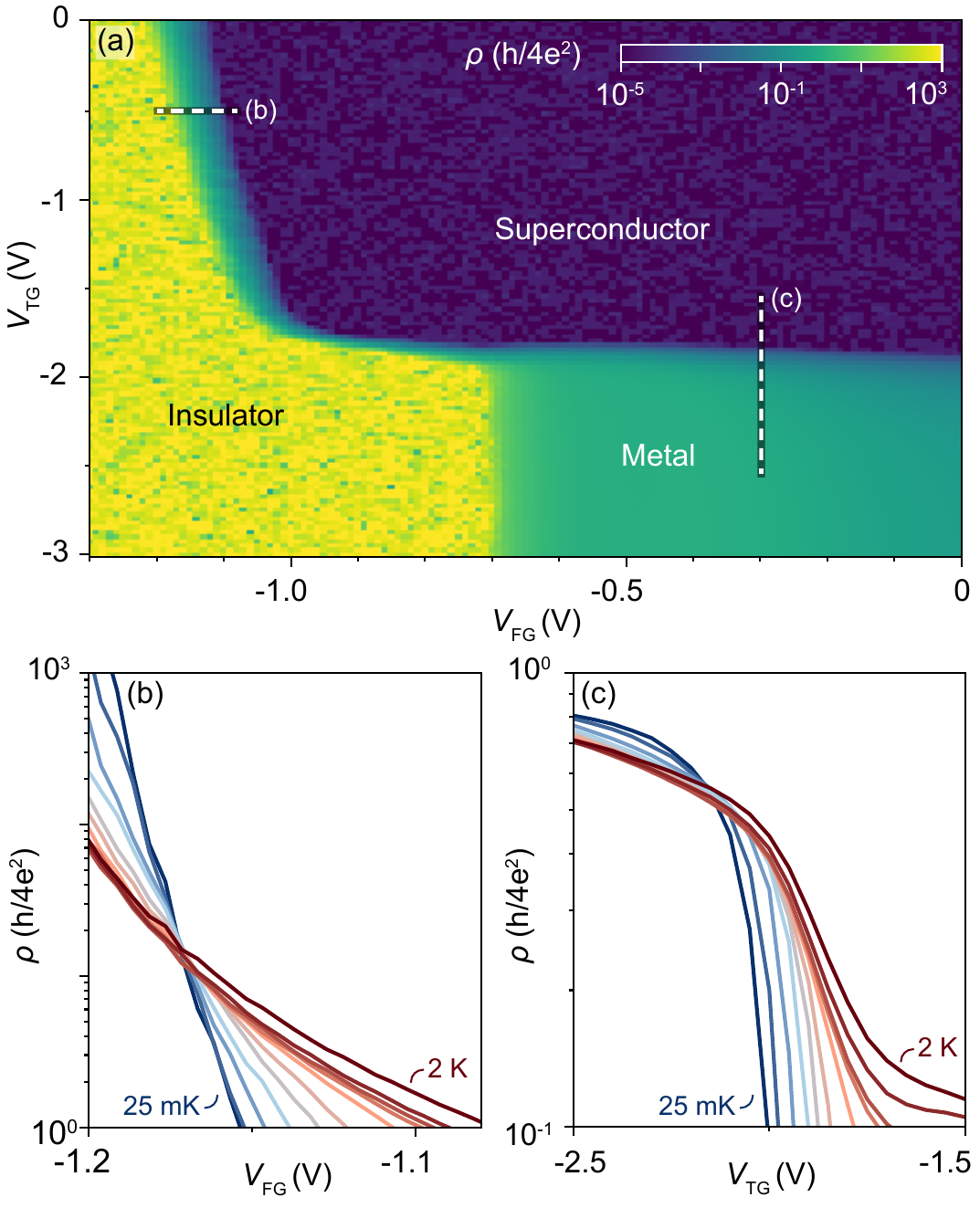}
    \caption{(a) Map of resistivity, $\rho$, measured for the main hybrid array as a function of top-gate, $V\sb{TG}$, and frame-gate, $V\sb{FG}$, voltages, showing the superconducting, insulating, and metallic states. 
    (b) Traces of $\rho$ as a function of $V\sb{FG}$ for different temperatures, $T$, and fixed $V\sb{TG}=-0.5$~V showing a crossing point around $h/4e^2$ indicating superconductor-insulator transition.
    (c) Similar to (b) but as a function of $V\sb{TG}$ at fixed $V\sb{FG}=-0.3$~V with a crossing point suggesting superconductor-metal transition.
    }
    \label{fig:2}
\end{figure}

We begin by characterizing the resistivity, $\rho$, of the hybrid array as a function of frame-gate, $V\sb{FG}$, and top-gate, $V\sb{TG}$, voltages at base temperature; see \cref{fig:2}(a).
For $V\sb{FG}=0$ and $V\sb{TG}=0$, the array is in a global superconducting state characterized by vanishing $\rho$.
This reflects coherent coupling between proximitized regions beneath neighboring islands via the Josephson effect.

Both gate voltages reduce the Josephson coupling between Al islands, but their effect on the global state of the device differs considerably.
When decreasing $V\sb{FG}$, we observe a transition into a highly resistive state as carriers in the junctions between islands are depleted, eventually disconnecting the islands.
The temperature evolution of $\rho(V_{\mathrm{FG}})$ at fixed $V\sb{TG}=-0.5$~V reveals a crossing point of all traces around $V\sb{FG}^* = -1.17$~V, where $\rho$ is roughly $h/4e^2$ and shows negligible $T$ dependence; see \cref{fig:2}(b). 
Below this point, $\rho$ increases with decreasing $T$, whereas above it, $\rho$ decreases.
This behavior is consistent with an SIT, comparable to previous observations in similar hybrid arrays~\cite{bottcher2018superconducting}.
In contrast, a negative $V\sb{TG}$ depletes the regions around the islands, weakening the proximity effect while maintaining conduction through the 2DEG under the frame gate.
At sufficiently low $V\sb{TG}$, the islands become effectively isolated and the array develops a finite $\rho$.
The corresponding $\rho(V_{\mathrm{TG}})$ traces also show a $T$-independent crossing point around $V\sb{TG}^*=-2.2$~V, though with weaker temperature dependence on the resistive side, confirming a superconductor-metal transition; see \cref{fig:2}(c).
Depending on the value of $V\sb{TG}$, sweeping $V\sb{FG}$ can induce either a superconductor-insulator or a metal-insulator transition.

\begin{figure}[t]
    \includegraphics[width=\linewidth]{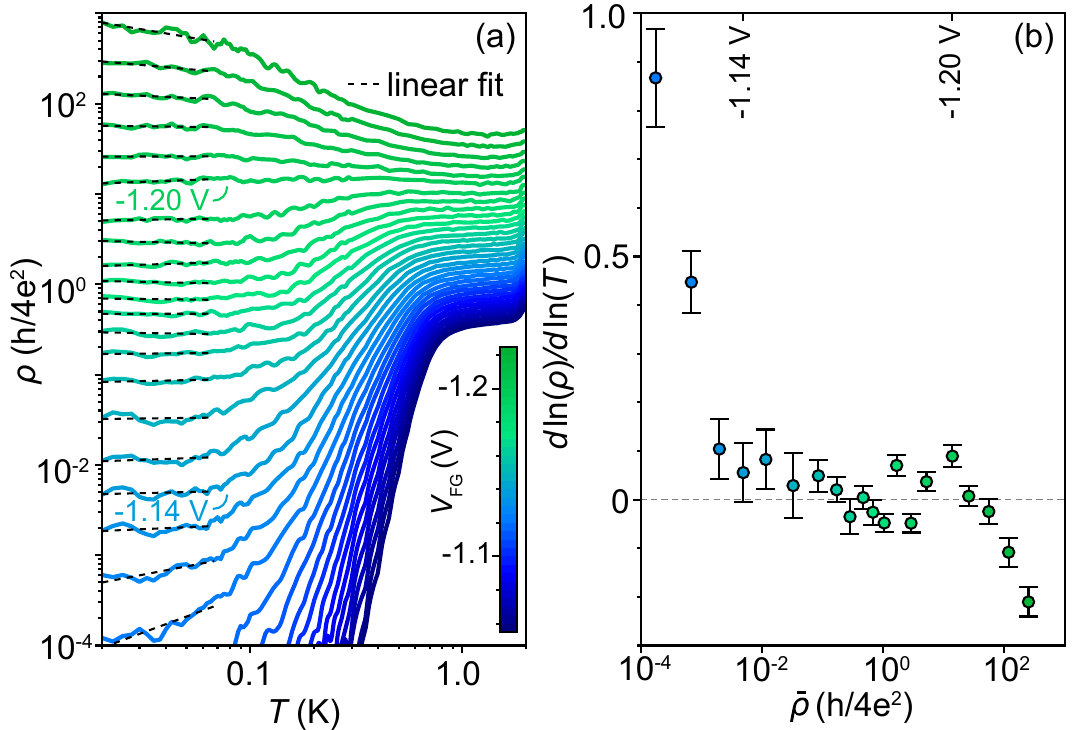}
    \caption{
    (a) Resistivity, $\rho$, as a function of temperature, $T$, measured for different frame-gate voltage, $V\sb{FG}$, values ranging from $-1.060$ to $-1.225$~V, at fixed $V\sb{TG}=0$.
    The dashed black lines are linear fits of $\ln(\rho)$ versus $\ln(T)$ up to $70$~mK.
    (b) Low-$T$ resistivity slope, $d \ln(\rho) / d \ln(T)$, taken from the fits in (a), plotted against the average resistivity, $\bar{\rho}$, in the same $T$ range.
    Error bars are the standard uncertainties of the fitting.
    }
    \label{fig:3}
\end{figure}

To better understand the nature of the voltage-tuned SIT, we investigate the temperature dependence of $\rho$ at different values of $V\sb{FG}$ while keeping $V\sb{TG} = 0$; see \cref{fig:3}(a).
For all $V\sb{FG}$ values, a small step in $\rho$ is visible around $T=1.8$~K, signifying the superconducting transition of the thin  Al islands~\cite{weerdenburg2023extreme}. 
As $T$ is decreased, traces at $V\sb{FG}\gtrsim -1.12$ V show a sharp drop in $\rho$ that eventually falls below the experimental sensitivity $(\lesssim 10^{-4}\,h/4e^2)$, confirming a global superconducting state.
For slightly more negative $V\sb{FG}\approx -1.14$~V, $\rho$ still drops initially but then saturates below $T\approx 100$~mK.
The saturation resistivity increases continuously with decreasing $V\sb{FG}$, spanning over three orders of magnitude (from roughly $ 10^{-3}$ to $10~h/4e^2$).
For $V\sb{FG}\lesssim-1.20$ V, $\rho$ starts to increase with decreasing $T$, indicating a crossover to the insulating state of the device.

To characterize the saturating regime at low temperature, we analyze the low-$T$ slope of $\ln(\rho)$ versus $\ln(T)$ for different $V\sb{FG}$.
For each trace, we perform a linear fit to the data below 70~mK, yielding $d\ln(\rho) / d\ln(T)$.
A positive (negative) slope indicates a superconducting (insulating) state.
The intermediate regime, with a suppressed slope but finite $\rho$, can be identified as the anomalous metal~\cite{kapitulnik2019colloquium}.
This can be illustrated by plotting $d\ln(\rho) / d\ln(T)$ as a function of  average resistivity, $\bar{\rho}$, in the fitted $T$ range; see \cref{fig:3}(b).
The slope is positive for $\bar{\rho}\lesssim10^{-2}~h/4e^2$ and negative for $\bar{\rho}\gtrsim10^2~h/4e^2$, consistent with superconducting and insulating states, respectively, as $T\to0$.
For $\bar{\rho}$ between $10^{-2}$ and $10~h/4e^2$, $d\ln(\rho) / d\ln(T)$ is roughly zero, indicating that the SIT is interrupted by the AM phase, spanning three orders of magnitude of $\bar{\rho}$. 

We note that $\rho$ at $V\sb{FG}=-1.20$~V remains roughly $h/4e^2$ throughout the measured $T$ range [\cref{fig:3}(a)].
For less negative $V\sb{FG}$, $\rho$ drops below its normal-state value---in some cases by several orders of magnitude---before saturating at low $T$.
For slightly more negative $V\sb{FG}$, the first couple of traces still saturate, but at values higher than their high-$T$ resistivity.
This behavior resembles a smooth crossover from a failed superconductor to a failed insulator within the anomalous metallic phase and may be attributed to quantum fluctuations~\cite{Couedo_2016_Dissipative, zhang2022anomalous, trugenberger2023gauge}. We note that a similar phenomenology has been observed in cuprate superconductors~\cite{li2019tuning}, although the connection between underlying microscopic mechanisms remains
unclear.

\begin{figure}[t]
    \includegraphics[width=\linewidth]{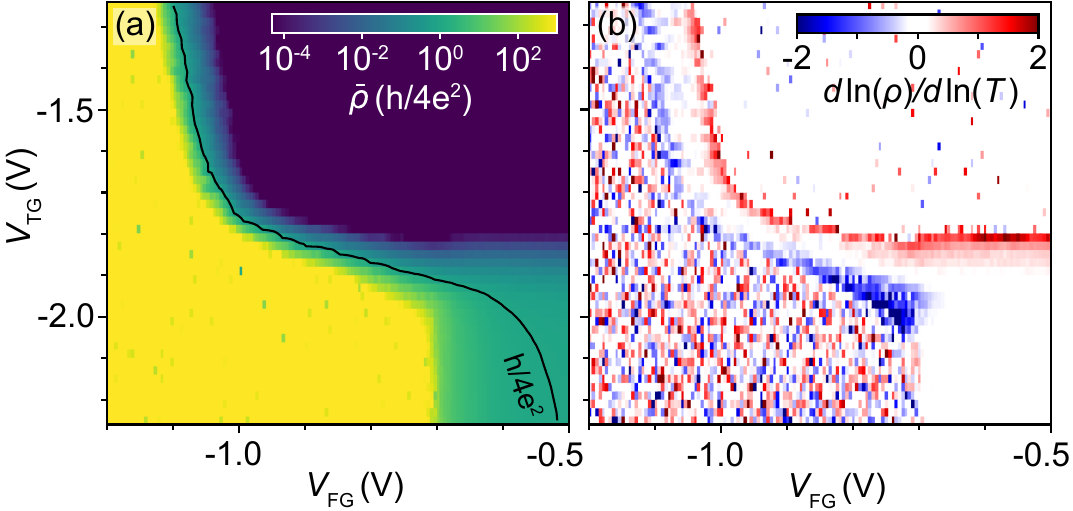}
    \caption{(a) Average resistivity, $\bar{\rho}$, measured as a function of top-gate, $V\sb{TG}$, and frame-gate, $V\sb{FG}$, voltages, showing an enlargement of the region where the superconducting, insulating, and metallic states meet. 
    The plot was constructed by averaging a set of $\rho$ maps over the temperature range from $T = 15$ to 55~mK, taken every $2.5$~mK.
    (b) Low-$T$ $\rho$ slope, $d \ln(\rho)/d \ln(T)$, obtained by linear fitting of $\ln(\rho)$ versus $\ln(T)$ for the same data as in (a)
    }
    \label{fig:4}
\end{figure}

The AM phase is observed as $V\sb{FG}$ reduces the carrier density in the junctions between the islands, driving the SIT, while the regions around the islands remain populated ($V\sb{TG} = 0$).
However, as mentioned earlier, depleting those regions can tune the system into a conventional metallic phase.
With this in mind, we investigate how the system evolves between these two metallic regimes.
Specifically, we focus on the narrow region that connects the intermediate-$\rho$ regime associated with the anomalous metal at less negative $V\sb{TG}$ to the broader metallic regime emerging at more negative $V\sb{TG}$; see \cref{fig:2}(a).
This connection is seen more clearly in a zoomed-in map of $\bar{\rho}(V\sb{FG}, V\sb{TG})$, averaged over $T = 15$ to 55~mK range taken in 2.5~mK steps; see \cref{fig:4}(a).

To further probe how the anomalous metal evolves into a conventional metal, we use the same data set to construct a map of the slope $d \ln(\rho)/d \ln(T)$; see \cref{fig:4}(b).
At the top-right corner of the map, global superconductivity yields vanishing resistance across the entire $T$ window, resulting in zero slope.
More negative gate voltages gradually suppress Josephson coupling, leading to finite resistivity with a positive slope [red regions in \cref{fig:4}(b)].
The left side of the map is dominated by noise, indicating that the islands become disconnected from the source and drain.
However, just before complete decoupling, the array exhibits insulating behavior with a negative slope [blue regions in \cref{fig:4}(b)].
The slope remains near zero between the superconducting and insulating boundaries, with no disruption along the narrow region as $V\sb{TG}$ decreases.
This continuity demonstrates a smooth evolution from anomalous to conventional metallic states, despite the distinct mechanisms driving the associated transitions.

\begin{figure}[t]
    \includegraphics[width=\linewidth]{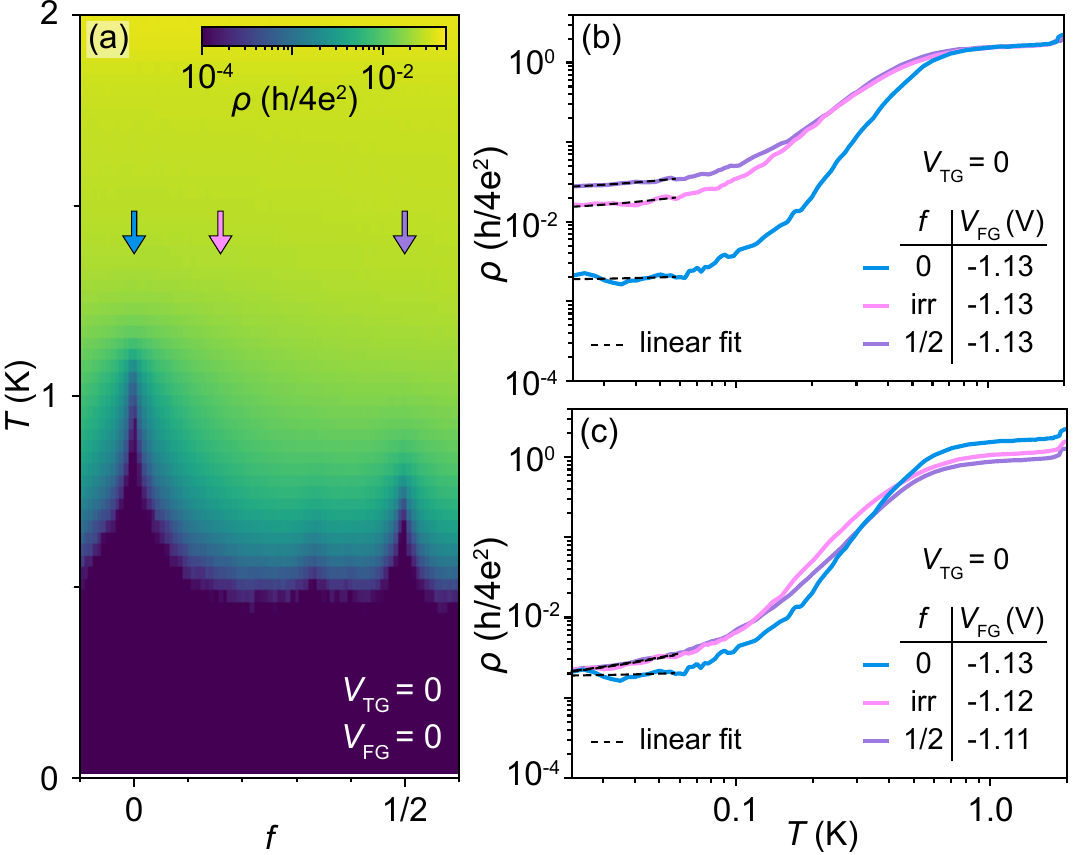}
    \caption{(a) Resistivity, $\rho$, measured as a function of temperature, $T$ and magnetic frustration, $f$, shows peaks of the superconducting transition temperature around commensurate frustration values, mostly pronounced around $f=0$ and $1/2$. 
    (b) Traces of $\rho$ as a function $T$, taken for three representative frustration values, $f=0$, irrational ($\approx 0.16$), and $f=1/2$, at fixed $V\sb{FG} = -1.13$~V.
    The normal-state resistivity matches for all the traces, but nonzero $f$ show higher $\rho$ at low $T$ with finite slope.
    (c) Similar to (b) but at $V\sb{FG}$ chosen such that $\rho$ at the base-$T$ match for all frustrations.
    }
    \label{fig:5}
\end{figure}

All data discussed so far were taken at magnetic field carefully tuned to zero (for more details, see Supplemental Material~\cite{Supplement}).
We now turn to investigate the effect of an out-of-plane magnetic field, $B$, which introduces a superconducting phase gradient between neighboring islands~\cite{teitel1983josephson}.
To this end, we map the resistivity as a function of $T$ and magnetic frustration, $f=\Phi/\Phi_0$, where $\Phi=BA$ is the magnetic flux through a unit cell of area $A$ and $\Phi_0 = h/2e$ is the superconducting flux quantum.
We find that the critical temperature, $T\sb{C}$, is maximal at $f=0$ and decays quickly as the array is frustrated; see \cref{fig:5}(a).
Additional peaks in $T\sb{C}$ appear at several other frustration values, most prominently at $f=1/2$, corresponding to commensurate vortex configurations. While the periodicity of the frustration pattern is geometry independent, we expect that narrower junctions or increased island spacing—both reducing the inter-island coupling—would lead to a stronger suppression of superconductivity with frustration.

To probe how frustration affects the AM phase, we compare $\rho(T)$ traces at $f=0$, irrational~($\approx 0.16$), and $1/2$ under two conditions.
First, we select traces with matching normal-state resistivity; see \cref{fig:5}(b).
At $f=0$, $\rho$ saturates at low $T$, characteristic of the anomalous metal.
In contrast, the traces with nonzero $f$ show a finite positive slope down to the base temperature, lying well above the $f=0$ trace.
Second, we keep the same $f=0$ trace and select traces with matching base-temperature resistivity; see \cref{fig:5}(c).
Here, the low-$T$ slope at nonzero $f$ remains finite, and their normal-state resistivity is lower than at $f=0$.

In both cases, the emergence of a small but finite positive slope at nonzero $f$ indicates that magnetic frustration destabilizes the low-temperature saturation of the AM phase. 
We attribute this behavior to thermally activated vortex motion transverse to the current flow, generating a longitudinal electric field \cite{behnia2016nernst}, which can introduce dissipation even before the onset of the Berezinskii–Kosterlitz–Thouless transition~\cite{bottcher2024berezinskii}. Additionally, the finite slope shows the system stays thermally sensitive at base temperature.

The fact that the resistance remains sensitive to $T$ down to the base temperature and evolves smoothly with gate voltage over a wide range suggests that the anomalous metallic phase observed at zero magnetic field is not a consequence of extrinsic effects, such as residual heating or environmental dissipation. Instead, the data are consistent with a scenario in which quantum fluctuations suppress long range phase coherence among proximitized superconducting islands. While the individual islands retain local Cooper pairing, the weakened inter-island coupling gives rise to superconducting phase fluctuations, leading to dissipation at low temperature. Possible sources of such fluctuations include microscopic inhomogeneities in Josephson coupling or charge offsets across the array, both of which may vary with gate voltage and enhance the effect of quantum fluctuations.

In summary, we have investigated voltage-controlled phase transitions in Josephson junction arrays patterned on an InAs/Al heterostructure.
Using a dual-gate geometry, we find a robust anomalous metallic phase interrupting the superconductor-insulator transition, which can be smoothly tuned into a conventional metallic state with fully decoupled superconducting islands.
At zero magnetic frustration, the anomalous metallic phase, characterized by vanishing low-temperature dependence, persists across more than three orders of magnitude in resistivity.
A finite frustration suppresses the resistivity saturation, indicating thermally activated vortex motion and confirming thermal sensitivity down to the base temperature. These results provide new experimental insights into the origin of the anomalous metallic phase.\\

\textit{Acknowledgments}---We thank S. Upadhyay for assistance with the device fabrication and L.~Banszerus, M.~R.~Lykkegaard, A.~Padley, and T.~Røhling for discussions. We acknowledge support from research grants (Projects No. 43951, No. 53097, and No. 50334) from VILLUM FONDEN, the Danish National Research Foundation, the European Research Council (Grant Agreement No. 856526), and the MEYS grant (Project No. LM202351).\\

\textit{Data availability}---
The data used to generate the figures in this work are available in Ref.~\cite{Sasmal_zenodo25}.

\bibliography{Literature_DOI.bib}

\end{document}


\title{Supplemental Material:\\Voltage-tuned anomalous-metal to metal transition in\\hybrid Josephson junction arrays}

\author{S.~Sasmal}
\affiliation{Center for Quantum Devices, Niels Bohr Institute, University of Copenhagen, 2100 Copenhagen, Denmark}
\author{M.~Efthymiou-Tsironi}
\affiliation{Center for Quantum Devices, Niels Bohr Institute, University of Copenhagen, 2100 Copenhagen, Denmark}
\affiliation{Department of Physics, Universitá del Salento, via Monteroni, 73100, Lecce, Italy}
\author{G.~Nagda}
\affiliation{Center for Quantum Devices, Niels Bohr Institute, University of Copenhagen, 2100 Copenhagen, Denmark}
\author{E.~Fugl}
\affiliation{Center for Quantum Devices, Niels Bohr Institute, University of Copenhagen, 2100 Copenhagen, Denmark}
\author{L.~L.~Olsen}
\affiliation{Center for Quantum Devices, Niels Bohr Institute, University of Copenhagen, 2100 Copenhagen, Denmark}
\author{F.~Krizek}
\affiliation{Center for Quantum Devices, Niels Bohr Institute, University of Copenhagen, 2100 Copenhagen, Denmark}
\affiliation{Institute of Physics, Czech Academy of Sciences, 162 00 Prague, Czech Republic}
\author{C.~M.~Marcus}
\affiliation{Center for Quantum Devices, Niels Bohr Institute, University of Copenhagen, 2100 Copenhagen, Denmark}
\affiliation{Materials Science and Engineering, and Department of Physics, University of Washington, Seattle WA 98195}
\author{S.~Vaitiek\.{e}nas}
\affiliation{Center for Quantum Devices, Niels Bohr Institute, University of Copenhagen, 2100 Copenhagen, Denmark}

\date{\today}

\maketitle

\section*{Sample Preparation}

The Josephson junction arrays were fabricated from a hybrid semiconductor-superconductor heterostructure realized using molecular beam epitaxy (MBE). The III–V semiconductor stack was grown on a semi-insulating Fe-doped InP substrate adapting the growth conditions from Ref.~\cite{cheah2023control}, which provides a comprehensive structural characterization of the heterostructure. It consists of an 8.5-nm InAs quantum well, encapsulated between a 6.3-nm In$_{0.75}$Ga$_{0.25}$As bottom barrier and a 13.4-nm top barrier of the same composition.
The heterostructure is capped with two monolayers of GaAs and a nominally 12-nm Al film grown and oxidized \textit{in situ} at approximately $-22^\circ$C.
The thin GaAs cap serves as an etch stop during device fabrication and as a diffusion barrier for the intermixing of In and Al.

\begin{figure}[b]
    \includegraphics[width=\linewidth]{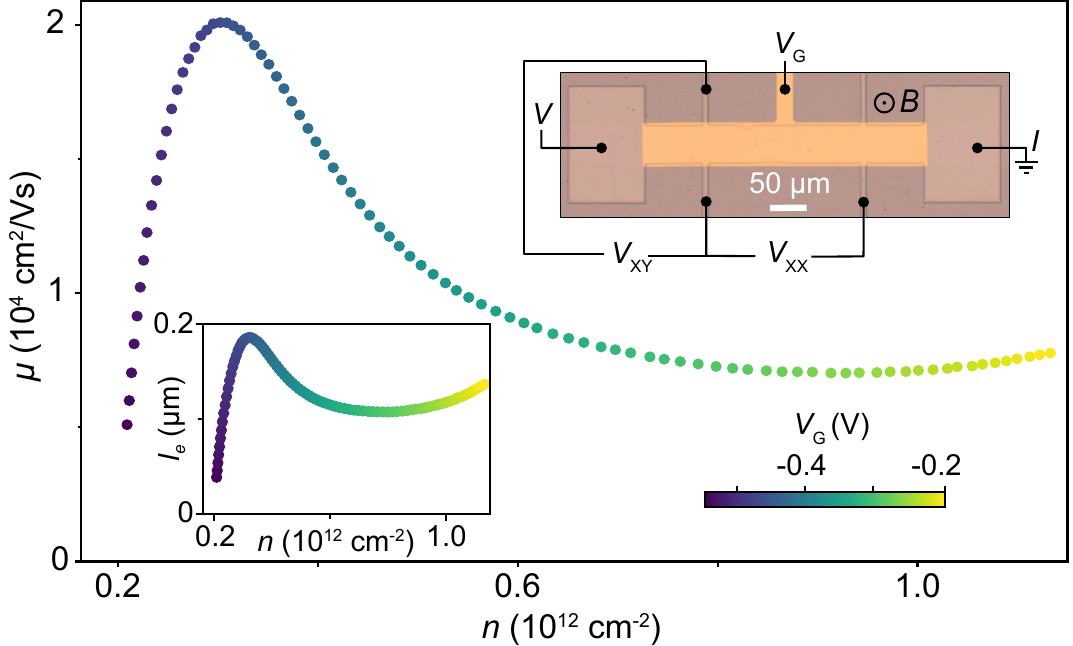}
    \caption{Charge carrier mobility, $\mu$, as a function of density, $n$, measured for a Hall bar without Al.
    The color represents the applied gate voltage, $V\sb{G}$.
    Left inset: Deduced mean free path, $l_e$, as a function of $n$.
    Right inset: Optical micrograph of the Hall-bar device and overlaid measurement setup.}
    \label{fig:s1}
\end{figure}

Device patterning was performed using standard electron beam lithography (Elionix 7000, 100~kV).
The mesa structures were defined using A4 PMMA resist and a chemical wet etch (220:55:3:3 H$_2$O:C$_6$H$_8$O$_7$:H$_3$PO$_4$:H$_2$O$_2$).
The normal metal Ti/Au (5/100~nm) ohmic pads were metallized (AJA International Inc., Orion) using A4 PMMA resist. 
The pattern of Al islands was selectively etched (Transcene, Aluminium Etchant Type D) at 50$^\circ$C for 5~s, using adhesion promoter (Allresist, AR 300-80~new) and A4 PMMA resist.
The chip was coated with a layer of HfOx (15~nm) gate dielectric using atomic layer deposition (Veeco, Savannah S100).
The frame gate was fabricated in two steps: a grid of Ti/Au (5/20~nm) was patterned on top of the mesa and then contacted by a thicker, mesa-climbing Ti/Au (15/380~nm) layer.
The second layer of HfOx (15~nm) was deposited before metallizing Ti/Au (15/380~nm) top gate.
Au bond wires were used to avoid magnetic field focusing.

To characterize the 2DEG, a top-gated Hall bar with Al layer removed was used to measure charge carrier mobility, $\mu$, and density, $n$. We find a peak mobility of $\mu\approx20, 000$~cm$^{2}$/V$\,$s at $n\approx0.3\times10^{12}$~cm$^{-2}$; see  \cref{fig:s1}. 

\begin{figure}[b]
    \includegraphics[width=\linewidth]{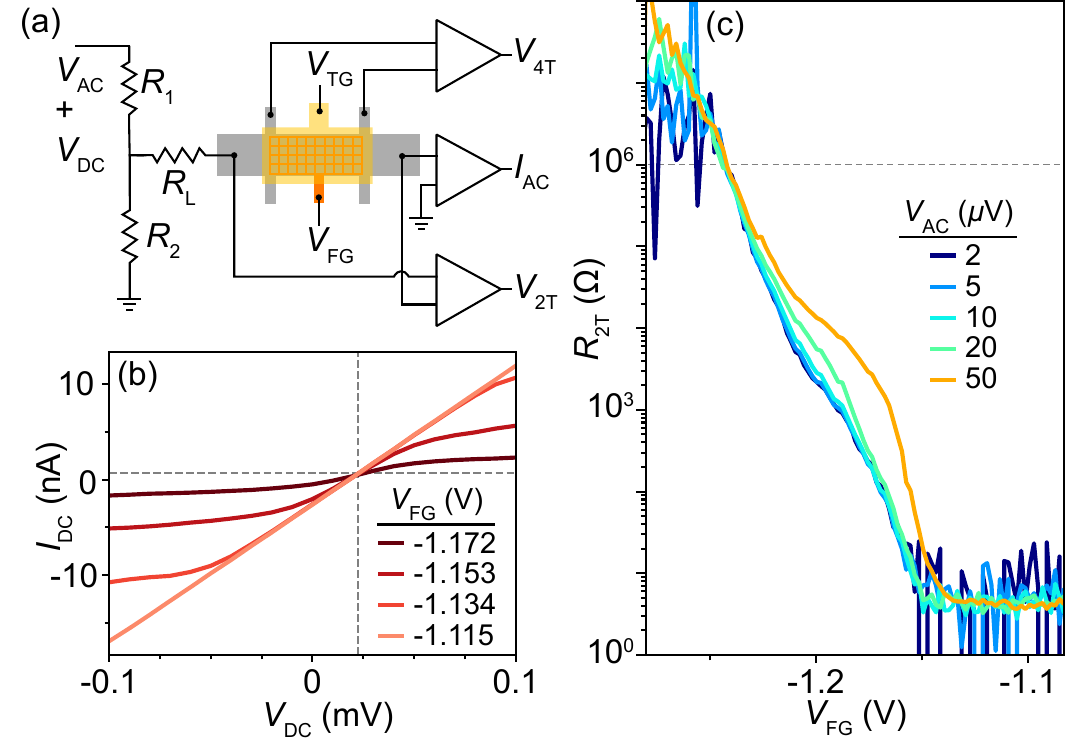}
    \caption{(a) Schematic of the measurement setup. 
    (b) Measured DC current, $I\sb{DC}$, as a function of applied DC voltage bias, $V\sb{DC}$, at different frame-gate voltages, $V\sb{FG}$, showing the offsets that were corrected before the resistivity measurements.
    (c) Two-terminal resistance, $R\sb{2T}$, as a function of $V\sb{FG}$ for different values of AC voltages bias, $V\sb{AC}$, measured at base temperature and top-gate voltage $V\sb{TG}=0$.}
    \label{fig:s2}
\end{figure}

Gate voltages in our devices primarily control the carrier density in the underlying InAs 2DEG. The
mobility peak indicates the full occupation of the lowest subband and minimal impurity scattering. The reduction of $\mu$ at lower $n$ reflects the increased effective disorder, while its decrease at higher $n$ is consistent with enhanced surface or inter-subband scattering. 
This non-monotonic behavior is also evident in the mean free path, $l_e = \mu \hbar \sqrt{2\pi n/|e|}$, which spans from several tens to hundreds of nanometers across the measured density range; see \cref{fig:s1}, inset. 
In the array devices, both the top gate, $V\sb{TG}$, and frame-gate, $V\sb{FG}$, voltages are expected to modulate the carrier density in a qualitatively similar manner, though with different lever arms due to different dielectric thicknesses.

\begin{figure}[t]
    \includegraphics[width=\linewidth]{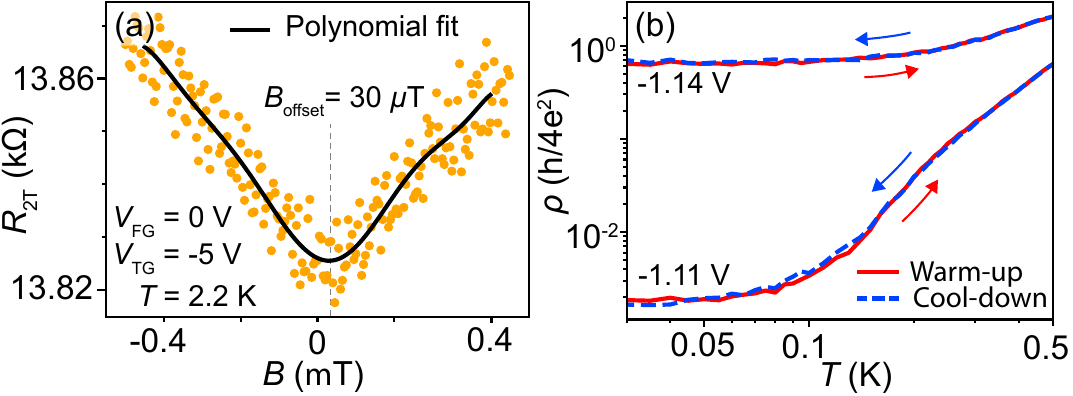}
    \caption{(a) Two-terminal resistance, $R\sb{2T}$, as a function of out-of-plane magnetic field, $B$, measured at $T=2.2$~K, $V\sb{FG}=0$, and $V\sb{TG}=-5$~V, showing a weak antilocalization dip.
    Solid black line is the polynomial fit to the data for estimating the magnetic field offset, $B\sb{offset}$.
    (b) Comparison of the resistivity, $\rho$, as a function of $T$, measured during warming (red) and cooling (blue) cycles for two representative $V\sb{FG}$ values. }
    \label{fig:s3}
\end{figure}

\section*{Measurement Setup}

Electrical measurements were carried out in a cryofree dilution refrigerator (Oxford Instruments, Triton 400) with a three-axis (1, 1, 6)~T vector magnet and a base temperature of 15 mK. 
The schematic measurement setup used to collect the data is shown in \cref{fig:s2}(a).
The AC and DC bias voltages were sourced using a lock-in amplifier (Stanford Research, SR860).
The bias was applied to the device via a home-built voltage divider ($R\sb{1}/R\sb{2}=10^{-3}$) and a load resistor $R\sb{L}=5$~k$\Omega$, in addition to roughly $1.6$~k$\Omega$ filter (line) resistances. 
The resulting AC current, $I\sb{AC}$, was transamplified using an $I$--$V$ converter (Basel Precision Instruments, SP983) and measured using a lock-in amplifier (Stanford Research, SR830).
In addition, differential 4-terminal, $V\sb{4T}$, and 2-terminal, $V\sb{2T}$, voltages were measured after being amplified (Stanford Research, SR560).

\section*{Setup Optimization}

To eliminate any effects due to nonlinear transport, we corrected the DC offset at base temperature by measuring the DC current, $I\sb{DC}$, in response to the applied DC voltage, $V\sb{DC}$, near the superconductor-insulator transition for several values of $V\sb{FG}$; see \cref{fig:s2}(b).
All traces intersect at  a single point, indicating the intrinsic voltage and current offsets (dashed lines in the figure), which were accounted for before the resistivity measurements.

To avoid AC bias-induced heating, we characterized the device response across the superconductor–insulator transition under varying AC bias voltages, $V\sb{AC}$. 
The two-terminal resistance, $R\sb{2T}$, was measured as a function of $V\sb{FG}$ at base temperature and $V\sb{TG} = 0$, for $V\sb{AC}$ ranging from 2 to $50~\mu$V; see \cref{fig:s2}(c). 
For $V\sb{AC}\lesssim10~\mu$V, the $R\sb{2T}$ curves overlap, indicating negligible heating.
The deviations at higher $V\sb{AC}$ signal the onset of bias-induced heating effects.
We therefore used $V\sb{AC}=10~\mu$V for low-resistance measurements to minimize heating and maximize signal-to-noise ratio.
For $R\sb{2T}>10^6$~$\Omega$, where the array resistance becomes comparable to the input impedance of the amplifier, we used $V\sb{AC}=50~\mu$V.

To correct the magnetic field offset, $B\sb{offset}$, originating from trapped flux in the superconducting magnet, we measured the magnetoresistance of the array at $T=2.2$~K, above the critical temperature of the islands.
The measured $R\sb{2T}$ as a function of out-of-plane magnetic field, $B$, shows a characteristic weak antilocalization (WAL) dip; see \cref{fig:s3}(a).
Assuming that the resistance minimum occurs where the net magnetic field is zero, we fit the data with a ninth-order polynomial to determine $B\sb{offset}$. 
Measurements were performed at $V\sb{FG} = 0$ and $V\sb{TG} = -5$~V to isolate the Al islands and enhance the WAL signal from the low-density 2DEG.
After correcting for $B\sb{offset}$, the device was cooled to base temperature to avoid vortex formation due to residual magnetic field.

\begin{figure}[t]
    \includegraphics[width=\linewidth]{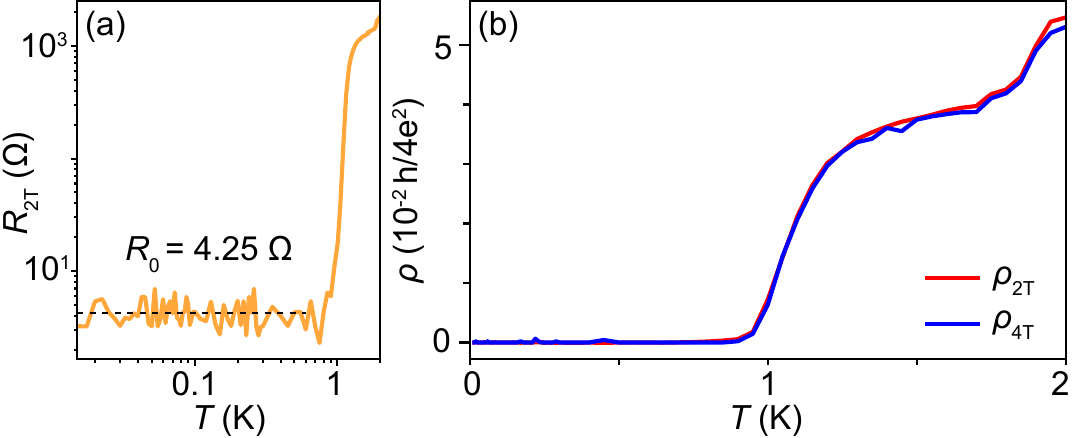}
    \caption{(a) Two-terminal resistance, $R\sb{2T}$, showing residual resistance of $R_0=4.25~\Omega$.
    (b) Comparison of sheet resistivity $\rho$ measured using (quasi-)two-terminal (2T) and four-terminal~(4T) configurations.
    Data were taken at $V\sb{FG} = 0$ and $V\sb{TG} = 0$.}
    \label{fig:s4}
\end{figure}

The resistivity, $\rho$, measured during both warming and cooling cycles at two representative frame-gate voltages, $V\sb{FG}$, shows good agreement; see \cref{fig:s3}(b).
This indicates good thermalization and minimal thermal hysteresis in the measurement setup.
Because cooling cycles often exhibit fluctuating temperature profiles due to nonuniform cooling rates, all $\rho(T)$ data presented in this work were acquired while increasing $T$.

These setup calibrations were essential to minimize residual heating, current noise, and voltage offsets that could obscure intrinsic low-temperature transport features. The measurements were carried out down to 15 mK--over two orders of magnitude lower than the superconducting transition temperature of the Al islands ($\sim 1.8$ K). In particular, reducing the AC excitation minimized Joule heating, while higher excitation was used only in the high-resistance regime ($>1$ M$\Omega$) to reduce current noise in the high-resistance regime. While quasiparticle excitations, in principle, produce an apparent saturation, we would expect such effects to be enhanced by magnetic field. Instead, the saturation is lifted under magnetic frustration (see Fig. 5 in the main text), suggesting that extrinsic mechanisms are unlikely to account for the observed behavior.

\begin{figure}[t]
    \includegraphics[width=\linewidth]{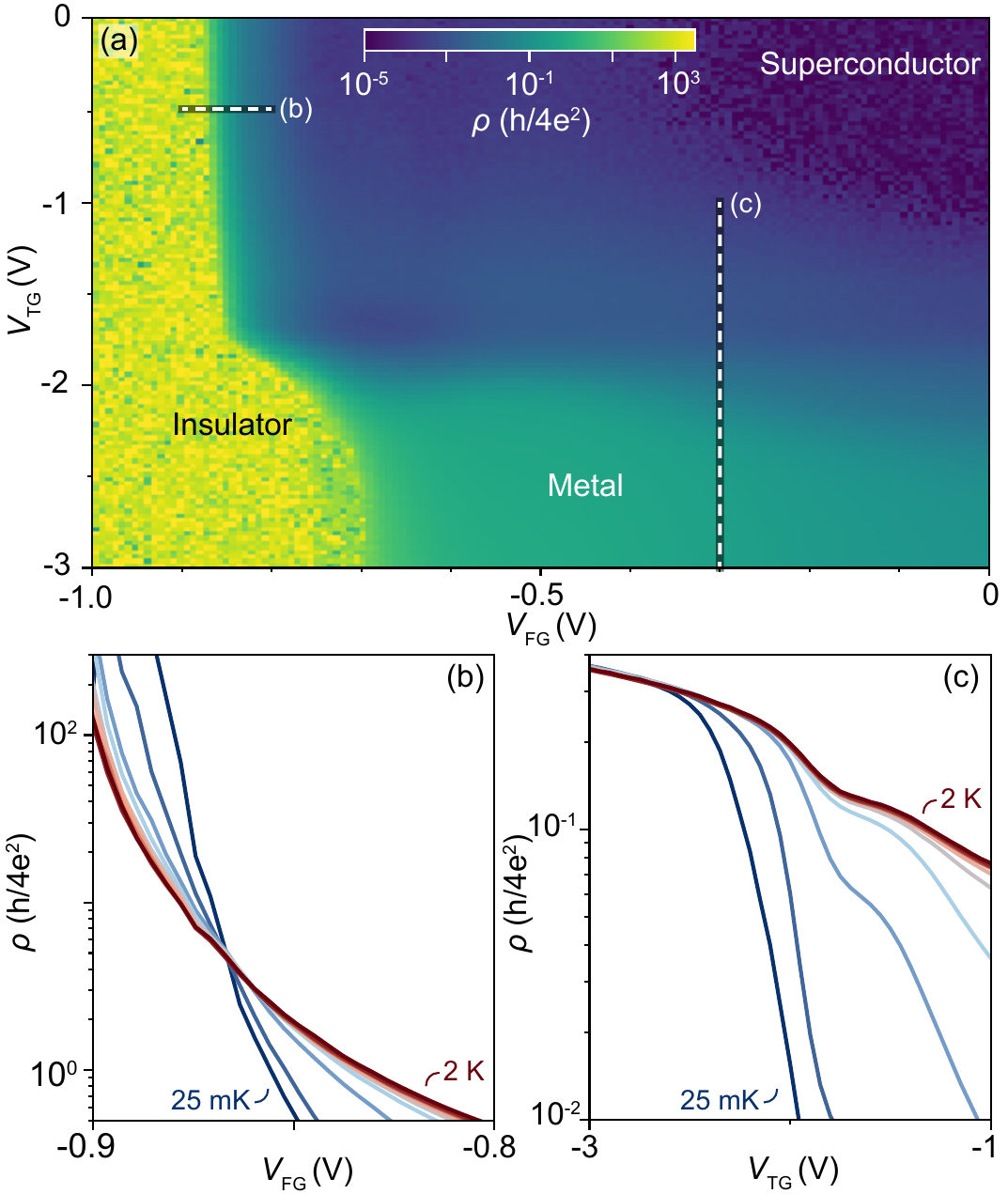}
    \caption{(a) Map of resistivity, $\rho$, measured for the first additional hybrid array a as a function of top-gate, $V\sb{TG}$, and frame-gate, $V\sb{FG}$, voltages, showing the superconducting, insulating, and metallic states. 
    (b) Traces of $\rho$ as a function of $V\sb{FG}$ for different temperatures, $T$, and fixed $V\sb{TG}=-0.5$~V showing a crossing point around $h/e^2$ indicating superconductor-insulator transition.
    (c) Similar to (b) but as a function of $V\sb{TG}$ at fixed $V\sb{FG}=-0.3$~V with a crossing point suggesting superconductor-metal transition.
    }
    \label{fig:s5}
\end{figure}

\section*{Resistivity Calibration}

Before the array undergoes the superconductor–insulator transition, the side probes intended for four-terminal (4T) voltage measurements are pinched off by the frame gate and become electrically disconnected from the device.
We therefore perform (quasi-)two-terminal (2T) measurements, using double-bonded source and drain leads, to determine the resistivity. 
This includes the transverse probes, which could otherwise be used to investigate the absence of a transverse signal--an additional signature of the anomalous metallic phase~\cite{breznay2017particle}.

The sheet resistivity is given by $\rho\sb{i} = (R_i - R\sb{0})\,W/L_i$, where $i$ refers to the measurement configuration (2T or 4T), $R_0$ is the residual resistance, $L_i$ is the effective length of the measured segment of the array, and $W$ is its width.
We determine the 2T $R_0$ from the average value of $V\sb{2T}/I\sb{AC}$ in the superconducting state---at $V\sb{FG} = 0$ and $V\sb{TG} = 0$, where the side-probes are not pinched off, and at low temperature where the resistance is flat; see \cref{fig:s4}(a).
For 4T measurements, $R\sb{0} = 0$ by definition.

A comparison between the resistivities measured in 2T and 4T configurations shows a good quantitative agreement across a wide range; see \cref{fig:s4}(b).
This consistency validates the (quasi-)two-terminal approach and enables reliable resistivity estimation throughout the full gate-voltage range.

\begin{figure}[t]
    \includegraphics[width=\linewidth]{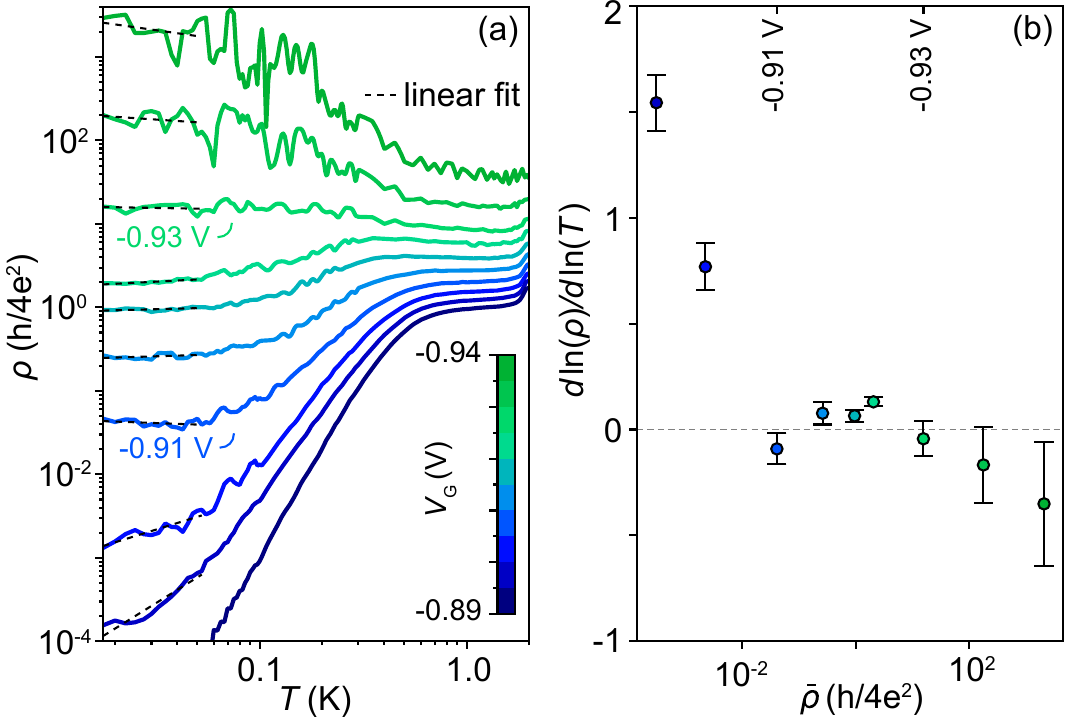}
    \caption{
    (a) Resistivity, $\rho$, of the second additional device as a function of temperature, $T$, measured for different combined gate voltage, $V\sb{G}\equiv V\sb{FG} = V\sb{TG}$.
    The dashed black lines are linear fits of $\ln(\rho)$ versus $\ln(T)$ up to $52.5$~mK.
    (b) Low-$T$ resistivity slope, $d \ln(\rho) / d \ln(T)$, taken from the fits in (a), plotted against the average resistivity, $\bar{\rho}$, in the same $T$ range.
    Error bars are the standard uncertainties of the fitting.
    }
    \label{fig:s6}
\end{figure}

\section*{Additional Measurements}

Two additional devices were investigated and showed similar results.
The first exhibits qualitatively the same
gate dependence of resistivity, $\rho$, as the main device, including the characteristic phase transitions; see \cref{fig:s5}.
It features a $498\times68$ array of $0.3~\mu$m wide square islands with 350~nm spacing and a 150~nm wide frame gate in between.
Despite these geometric differences--smaller islands and tighter spacing--the overall transport characteristics closely match those of the main device, supporting the generality of the observed phenomena.

The second additional device consists of a $264\times32$ array of $1~\mu$m sized square islands with 350~nm spacing and a 150~nm wide frame gate.
In this device, the top and frame gates were electrically shorted, and both were biased at the same potential, $V\sb{G} \equiv V\sb{FG} = V\sb{TG}$.
Despite this simplified gating configuration, the array exhibits a superconductor–insulator transition as a function of $V\sb{G}$, with an extended anomalous metallic phase interrupting the two phases—similar to the main hybrid array; see \cref{fig:s6}.


\bibliography{Literature_DOI.bib}